          [2004/06/22 1.2 (KOF); 2001/04/25 1.1 (PWD)]

\documentclass[a4paper,twocolumn]{Gaia2004} 
\usepackage{times}      
\usepackage{epsfig}     
\usepackage{natbib}     
\title{Cepheid Period-Luminosity Relations: Galactic vs. LMC and the Results from {\it t}-Test.}

\author{Ngeow, C. \& Kanbur, S.}
\affil{University of Massachusetts (Amherst, USA)}

\bibpunct{(}{)}{;}{a}{}{,}  

\begin{document}

\keywords{Cepheid; Distance Scale}

\maketitle

\begin{abstract}
  
The period-luminosity (PL) relation for Galactic Cepheids is derived with recent independent distance measurements taken from literature. Our PL relation confirms the work of Tammann et al. (2003), which showed that the Galactic Cepheids follow a different PL relation to their LMC counterparts. Our results also show that the slope of the Galactic PL relation is inconsistent with the LMC slope with more than 95\% confidence. The details of this work can be found in Ngeow \& Kanbur (2004).

\end{abstract}

\section{Introduction}

The Cepheid period-luminosity (PL) relation is important in deriving the distance to nearby ($<30Mpc$) galaxies. This relation takes a very simple form: $M_{\lambda}=a_{\lambda}\log(P)+b_{\lambda}$, where $M$, $a$ and $b$ are the absolute magnitudes, slope and zero-point (ZP) in bandpass $\lambda$, respectively, and $P$ is the pulsational period in days. Both of the slopes and ZP can be obtained either from the theoretical predictions or calibrated from the observations.

In most cases, the Cepheid PL relation is assumed to be {\bf universal}, i.e. the slope of the PL relation remains fixed, and the distance mudulus to a target galaxy is found by obtaing the {\it difference} of the ZPs from the calibrated and the fitted PL relation to the Cepheids in target galaxy. The most-widely used PL relations are calibrated with the large number and well observed Cepheids in the Large Magellanic Cloud (LMC), by adopting a distance modulus of $18.5\pm0.1mag.$ to the LMC \citep[see, e.g., ][]{fre01}. 

There are numerous studies on the metallicity dependency of the PL relation \citep[see, for example, ][]{sto88,fre90,koc97,sas97,ken98,cap00,fre01,kan03,gro04,rom04,sak04,sto04}. Most of these studies suggest that the metallicity dependency enters the PL relation through the ZP, while the slope is still univeral and fixed by the LMC Cepheids. Therefore, a metallicity correction term is commonly applied to derive the Cepheid distances to nearby galaxies that have a different metallicity environment to that of the LMC \citep[for examples, as in ][]{fre01,kan03,leo03}. However, a recent study by \citet{tam03} shows that the Galactic PL relation is {\it steeper} than the LMC PL relation, which suggests that the {\it slope} of the PL relation may also depends on the metallicity. Recall that the values of the metallicity for Galaxy (with $Z=0.020$) and LMC (with $Z=0.008$) are: 

\begin{itemize}
\item  $12+\log[O/H]_{LMC} = 8.50\pm0.08\ dex$, \citep{fer00}.
\item  $12+\log[O/H]_{GAL} = 8.87\pm0.07\ dex$, \citep{gre96}.
\end{itemize}

Therefore, the metallicity dependency on the slope of the PL relation can be tested by comparing the calibrated Galactic and the LMC PL relations. The difference between this study and the work of \citet{tam03} are:

\begin{enumerate}
\item  We include some additional Galactic Cepheids that have independent distance measurements from recent literature.
\item  We took the weighted average from different distance measurements for a given Cepheid as the final adopted distance.
\item  We use a statistical test to examine the consistency of the slopes for the Galactic and LMC PL relations. 
\end{enumerate}

The details of this work is presented in \citet{nge04}.

\section{Calibrating the Galactic Period-Luminosity relation}

In order to calibrate the Galactic PL relation, the distances to the Galactic Cepheids need to be known via independent measurements (i.e., not depend on the {\it assumed} PL relation). There are a few ways to measure the independent distances to the Galactic Cepheids:

\begin{enumerate}
\item[I.]  {\bf Open Cluster Distance:} If the membership of a Cepheid in the open cluster can be verified \citep{tur88}, then the distance of this Cepheid is equal to the distance of the open cluster. The distance of the open cluster can be found via the main-sequence fitting method \citep[see, e.g.,][]{tur98,fea99,tur02,hoy03}. 

\item[II.]  {\bf Baade-Wesslink Method:} The Baade-Wesslink (BS) method involves the comparison of the angular size of the Cepheid and its radius to obtain the distance. The radius of the Cepheid is obtained via the integration of the velocity curve:

  \[ R(t) \propto \int V_r(t)dt. \]

  There are two ways of obtaining the angular size of the Cepheids, as follows:

  \begin{enumerate}
  \item  Barnes-Evans (BE) surface brightness techniques -- The BE surface brightness techniques are first introduced by \citet{bar76} and frequently applied with the BW method to obtain the Cepheid distances. The idea behind the BE surface brightness techniques is simple: starting from the Stefan-Boltzmann law, $L=4\pi R^2\sigma T^4$, dividing by $4\pi D^2$ (where $D$ is the distance) on both sides, we obtain the expression for the angular size ($\theta$) after the $\log(L)$-$V_o$ and $\log(T)$-$(V-K)_o$ conversions:

    \[  \log(\theta) = \alpha V_o + \beta (V-K)_o + \gamma. \]
  
    The coefficients in this expression can be obtained with the independent calibration, such as using supergiant stars \citep{fou97}. Detailed applications of this method can be found, e.g., in \citet{gie93,gie97,gie98}.

  \item  Optical interferometry -- By using optical interferometry, the angular sizes for some large and/or nearby Cepheids can be obtained from the measured visibility \citep{lan02,nor02,ker04}. 
  \end{enumerate}

\item[III.]  {\bf Parallax/Astrometry:} The distances to Galactic Cepheids can also be obtained from the trigonometrical parallax with {\it Hipparcos} or the astrometry measurement from the {\it HST}. However, after converting the parallax to distance modulus, the error bars for the {\it Hipparcos} measurements are large, as can be seen in figure 2 of \citet{mad98}. Therefore, the {\it Hipparcos} measurements are not used in this study. For the {\it HST} astrometry measurements, currently there is only one Cephied, $\delta$ CEP, with the distance measurement from {\it HST} astrometry \citep{ben02}.
\end{enumerate}

\begin{figure}[h]
  \begin{center}
    \leavevmode
    \centerline{\epsfig{file=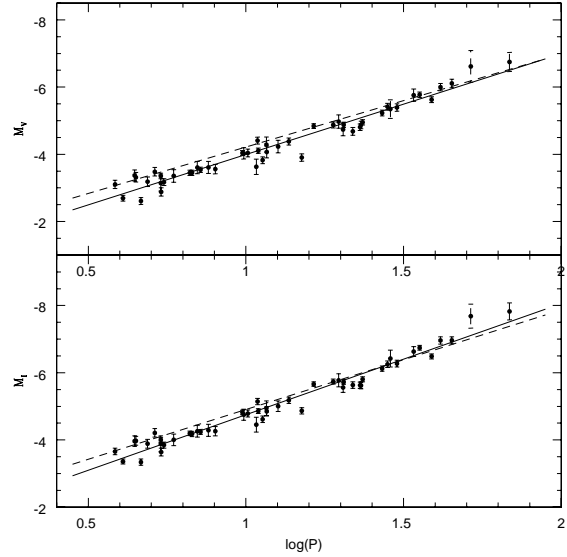,width=8.0cm}}
  \end{center}
  \caption{The Galactic PL relations in {\it V}- and {\it I}-band. The solid lines are the linear least squares fit to the data. The dashed lines are the LMC PL relations, for comparison.}
  \label{fig}
\end{figure}

We collect the available distance measurements with these defferent techniques from the literature, and took the weighted average for those Cepheids with different independent distance measurements. The final list contains 50 Galactic Cepheids, and the Galactic PL relations are constructed with the data of these Cepheids, as shown in Figure \ref{fig}. The list of these Cepheids can be found in \citet{nge04}.

\section{Comparison of the Galactic and the LMC PL relations}

\begin{figure}[h]
  \begin{center}
    \leavevmode
    \centerline{\epsfig{file=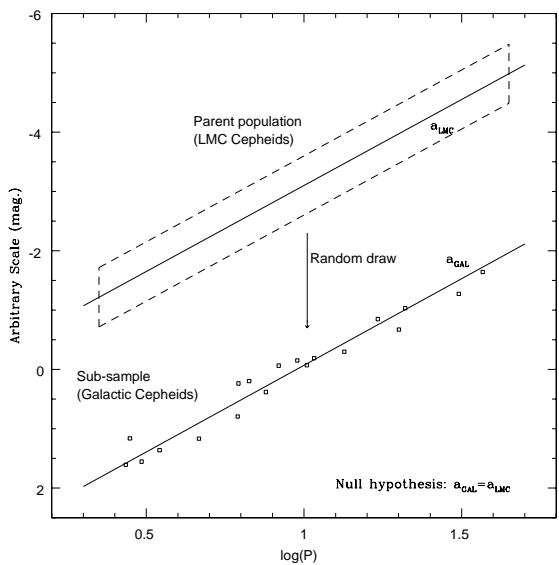,width=8.0cm}}
  \end{center}
  \caption{Illustration of the idea behind the $t$-test for the slopes of the Galactic PL relation.}
  \label{fig2}
\end{figure}

The slopes of the resulting Galactic {\it V}- and {\it I}-band PL relations from Figure \ref{fig} are given in Table \ref{tab1}. These slopes are consistent with the Galactic slopes found in \citet{tam03}. In this table, we also include the slopes of the LMC PL relations used in the $H_0$ Key Project \citep[$H_0$KP, ][]{fre01}. These LMC PL relations are obtaind with $\sim650$ Cepheids from the Optical Gravitational Lensing Experiment \citep{uda99}. As can be seen from the table, the slopes of the Galactic PL relations are more than $2\sigma$ away from their LMC counterparts. Are the slopes form Galactic Cepheids consistent with the slopes of the LMC PL relation? We can use the {\it t}-statistical test to answer this question \citep[see, for example, ][]{zwi00}. Assume that the Galactic data are (randomly) drawn from the same parent population, which is consist of the LMC data, then the distribution of the Galactic slopes should follow the {\it t}-distribution under the null hypothesis that $a_{GAL}=a_{LMC}$ (see Figure \ref{fig2} for illustration). By calculating the corresponding $t$ values, we can obtain the $p$-values (or the probability) for the null hypothesis. The results are:

\begin{description}
\item  For {\it V}-band slope, $p(V)=0.017$.
\item  For {\it I}-band slope, $p(I)=0.001$.
\end{description}
  
The small $p$-values indicate that the null hypothesis can be rejected with more than 95\% confidence level. Therefore the slopes of the {\it V}- and {\it I}-band Galactic PL relation are not consistent with their LMC counterparts. These statistical results strongly suggest that the PL relation is {\bf not universal}, at least in the galaxies with metallicity that comparable to the Galactic or LMC values.

\begin{table}[htb]
  \caption{Comparison of the Galactic and the LMC PL relations}
  \label{tab1}
  \begin{center}
    \leavevmode
    \begin{tabular}[h]{lcc} \hline
      PL relation & $a_V$ &  $a_I$ \\ 
      \hline \hline  
      GAL (Here) & $-2.999\pm0.097$ & $-3.303\pm0.094$  \\
      LMC ($H_0$KP) & $-2.760\pm0.030$ & $-2.962\pm0.020$  \\ 
      \hline    
    \end{tabular}
  \end{center}
\end{table}

\section{Conclusion \& Discussion}

In this study, we compare the slopes of the Galactic and the LMC PL relations with a $t$-test, and the results strongly suggest that the PL relation is not universal. However the current calibrated Galactic PL relations are based on 50 Cepheids, and there are about 500 Cepheids in our Galaxy (see the Galactic database maintained by \citealt{fer95}\footnote{\tt www.astro.utoronto.ca/DDO/research/cepheids/}). Therefore the astrometric and photometric measurements from the {\it Gaia} mission (see, e.g., the separate papers by Eyer, Mignard and Perryman in this Symposium) to the Galactic Cepheids can greatly improve the calibration of the Galactic PL relation and provide a larger statistical sample for the $t$-test. The importances of having the accurately calibrated Galactic PL relation are two-fold:

\begin{enumerate}
\item  Distance scale study: The distances to the high metallicity galaxies, either in the $H_0$KP or in the future observations, should be obtained with the Galactic PL relations. The attempts of using the Galactic PL relations to calibrate the Cepheid distance to the $H_0$KP target galaxies can be found in \citet{kan03}.
\item  Constrain on the pulsational/evolutionary models: The slopes from the theoretical Galactic PL relations are much {\it shallower} \citep{fio02} than the observed slopes as presented here, therefore the improvements for the theoretical Galactic PL relation is desired.
  \end{enumerate} 

Finally, it is worth while to point out that the study by \citet{hea04}, who applied the Bayesian analysis to the {\it Hipparcos} data, has found that the Galactic slope is much shallower than the slopes found in this study or even in the LMC. Therefore, the results from the {\it Gaia} mission is clearly desirable to solve this discrepancy.
 
\section*{Acknowledgments}

CN would like to acknowledge the support from the Organizing Committee of {\it Gaia} Symposium, who provides the grants for the registration fee and the accomodation; and the American Astronomical Society and the National Science Fundation for providing the international travel grants to cover the airfare.

\end{document}